\documentclass[11pt,twoside]{article} 
\usepackage{asp2004}
\usepackage{natbib}
\usepackage{epsf}
\usepackage{psfig}
\usepackage{lscape} 
\usepackage{verbatim}
\begin{document}
\title{Metal Abundances in Hot DA White Dwarfs}
\author{S.~Schuh,$^1$ M. A.~Barstow,$^2$ and S.~Dreizler$^1$}
\affil{$^1$Institut f\"ur Astrophysik, Universit\"at G\"ottingen,
  Friedrich-Hund-Platz\,1, D--37077~G\"ottingen, Germany\\
  $^2$Department of Physics and Astronomy, University of Leicester, 
  University Road, Leicester, LE1~7RH, United Kingdom}
\begin{abstract}
  We compare measured element abundances in hot DA white dwarfs from
  UV observations to predictions from our self-consistent non-LTE
  model atmosphere diffusion calculations.
\end{abstract}
\section{Introduction}
Due to their high surface gravity, the atmospheres of white
dwarfs exhibit a quasi-monoelemental composition: a large fraction of
all heavy elements has disappeared from the outer layers due to
gravitational sedimentation. Traces of metals may however be sustained
by radiative levitation. The radiative acceleration is
exerted on trace elements by a non-LTE radiation field through the
element's local opacity and therefore can vary strongly with depth,
which results in a chemically stratified atmospheric stucture.
\par
In an attempt to describe the chemically stratified atmospheres of hot
white dwarfs, modifications to our model atmosphere program have been
implemented to allow the self-consistent prediction of depth dependent
abundance profiles.
\par
Balancing the radiative acceleration and the effective gravitational
acceleration (including the effects of the electrical field that
builds up through diffusion of electrons) yields an equilibrium
condition for each atomic species. Its solution yields equilibrium
abundances. We present theoretical predictions from our
model grid in comparison to previous calculations as well as to
abundances measured from IUE and HST spectra.
\begin{figure}
  \plotone{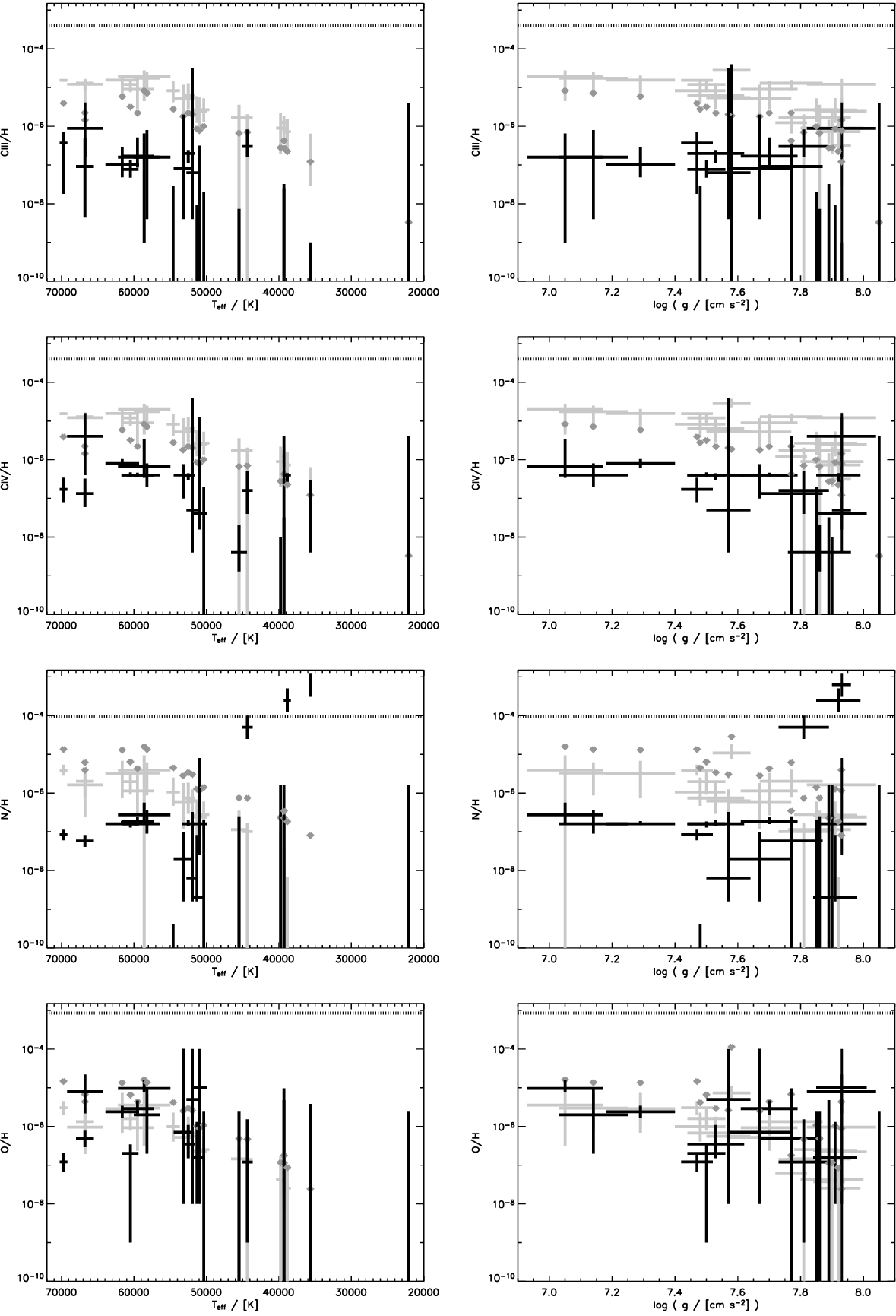}
\end{figure}
\begin{figure}
  \plotone{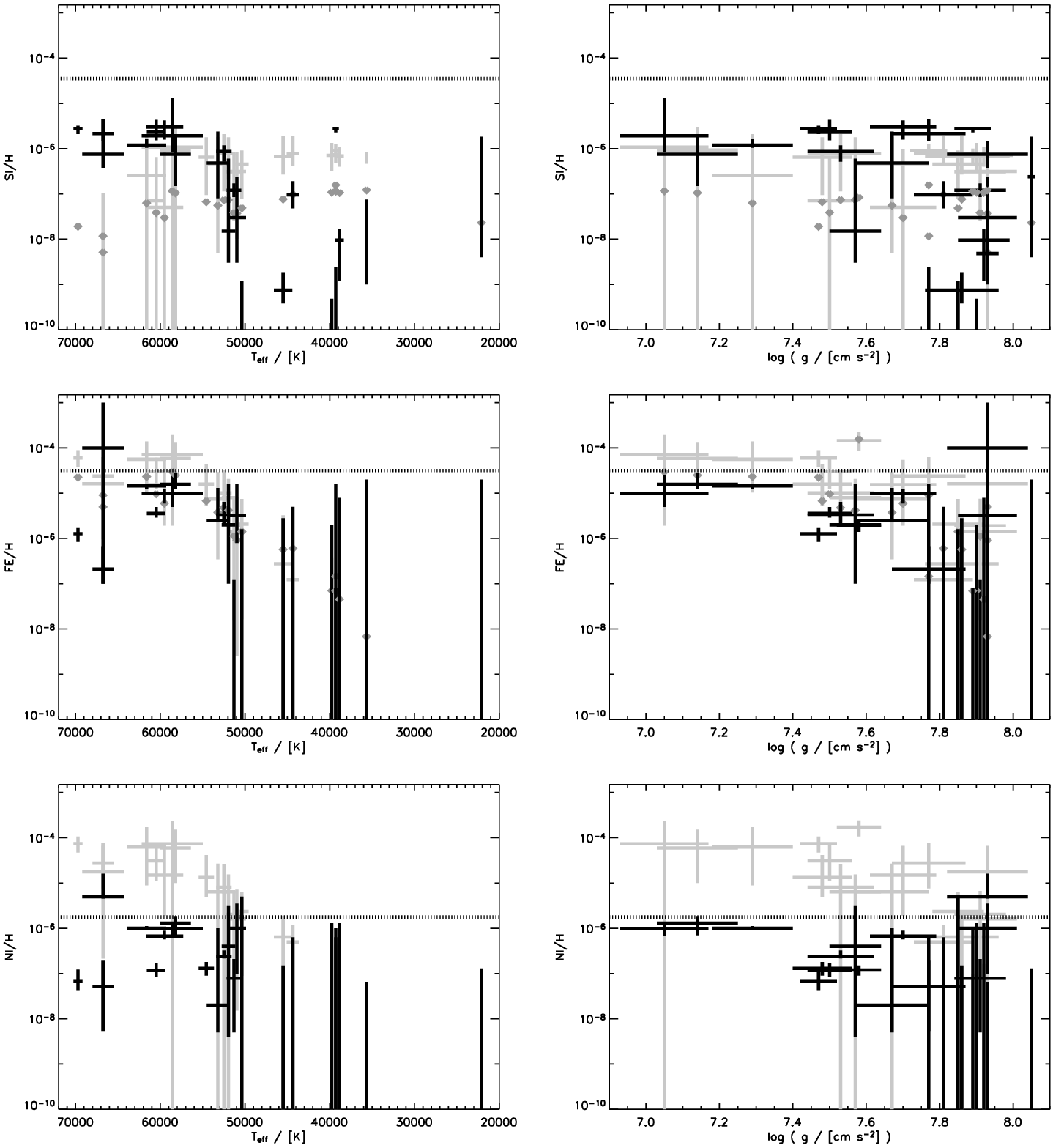}
  \vspace{12mm}
  \caption{(continued from previous page)
    \quad
    Black symbols: abundances measured by
    \citet[B03]{barstow:03} from observations using homogeneous
    models; errors in $T_{\rm eff}$ and abundances are
    the formal fitting errors taken from B03.
    \quad
    Grey symbols: diffusion model
    predictions interpolated for $T_{\rm eff}$ and $\log{g}$
    for each object as given in B03; $T_{\rm eff}$ errors 
    as above. Error bars for abundances obtained by
    evaluating the abundance variation due to the given
    uncertainties in $T_{\rm eff}$ and $\log{g}$.
    \quad
    Small dark grey symbols: the same for diffusion models by
    \citet{chayer:95b} but without errors assigned. 
    \quad
    Dotted line: cosmic abundance of the element.
  }
  \label{fig:schuhposter1}
\end{figure}
\section{Model Grid and Comparison to Models by \citet{chayer:95a,chayer:95b}}
Self-consistent NLTE diffusion models are available in
a $T_{\rm eff}$ range from $38\,000$\,K to $71\,000$\,K
for $\log{g}$ between $7.2$ and $8.4$ \citep{2002A&A...382..164S}.
From the full depth-dependent abundance stratification patterns, only the
$\tau_{\rm ross}=\frac{2}{3}$ values are used in
Fig.\,\ref{fig:schuhposter1}. Full abundance tables, as well as
high-resolution optical, UV, and far-UV spectra (not shown
here) are also 	available.
\par
Equilibrium abundances for $\tau_{\rm ross}=\frac{2}{3}$ published by
\citet[LTE, no iteration; no data for nickel]{chayer:95a,chayer:95b}
are shown for comparison using small dark grey symbols. In terms of
''evolution'' of diffusion codes, the systematic effects from one
generation to the next are still considerable.
\section{Comparison to Measurements by \citet{barstow:03}}
For carbon, both \ion{C}{\sc iii} and \ion{C}{\sc iv} are consistently
over-predicted. Theory and observations follow the same general
temperature dependency but although theory runs in parallel to the
observations its level is too high.
\par
Nitrogen, although not as clearly as for carbon, is also
over-predicted at higher temperatures; towards lower
temperatures, the models follow the lower branch of the
observed dichotomy.
For an earlier discussion of one of the nitrogen-rich objects
(RE\,J1032$+$535) see \citealt{holberg:99b} (besides B03).
\par
Oxygen generally follows the trend with temperature well, and the
predictions are mostly consistent with the observational error bars. Compare
also the successful application presented by \citet{2003whdw.conf..127C}. 
\par
Silicon, in contrast to {C} and {N},
shows opposite gradients with respect to temperature in theory and
observation, effectively leading to a similarly good agreement as for
{O} in the cross-over area but to under-predictions at
higher temperatures and over-predictions below $\approx 50\,000$\,K.
The only exception is {GD\,394}, considered to
show an anomalously high silicon abundance, for which the
absolute value is approximately reproduced by the models.
\par
Iron is consistent with observations on a star by star basis over
the full temperature range. This is in
agreement with EUV observations (where {Fe} is the
most important source of opacity) which can successfully be
reproduced with the stratified diffusion models.
On average across all stars however, the predicted iron abundance
remains about a factor 2 higher than observed, a value of
the order of the systematic error expected for the models.
\par
Nickel should behave similarly to {Fe} within the framework of
radiative levitation theory, which it effectively does, but this
predicted behaviour is in disagreement with observations. The observed
{Ni} is well below the prediction so that instead of
$\frac{{Fe}}{{Ni}}\approx1$, {Fe} and {Ni} are present in a roughly
cosmic ratio (as already stated in B03).
\par
With very few exceptions (objects belonging to the upper
branch of nitrogen dichotomy, iron and nickel in
\mbox{PG\,1342+444}), all photospheric abundances seem to respect
cosmic abundances as an upper limit. Given that equilibrium
radiative levitation theory ignores any evolutionary
constraints, this may be indicative of a ''reservoir problem''
(no unlimited supply of all elements available)
in the real stars.
\acknowledgements{Thanks to K.~Werner, I.~Hubeny and P.~Chayer for
  helpful discussions. Funding included DFG grants
  DR~281/13-1+2, travel grants by the Schuler
  Stiftung, the U. of Leicester and the conference organizers.
}
\end{document}